\documentstyle[aps,preprint,eqsecnum]{revtex}
\begin{document}

\title{  Non-locality and Medium Effects in the Exclusive
 Photoproduction of
$\eta$ Mesons on  Nuclei\thanks{
            Work supported in part by the Natural Sciences and
            Engineering Research Council of Canada} }

\author{\bf  M. Hedayati-Poor$^{a,c}$, S. Bayegan$^b$ and H.S. Sherif$^a$ \\
(a)Department of Physics, University of Alberta \\
Edmonton,  Alberta, Canada T6G 2J1\\
(b)Department of Physics, University of Tehran, Iran \\
(c)Department of Physics, University of Arak, Iran}

\date{\today}

\maketitle

\begin{abstract}
A relativistic model for the quasifree exclusive photoproduction of $\eta$ 
mesons on nuclei is extended to include both non-local and medium effects. 
The reaction is assumed to proceed via the dominant contribution of the  
S$_{11}$(1535) resonance. 
The complicated integrals resulting from the non-locality are simplified using 
a modified version of a method given by Cooper and Maxwell. 
The non-locality effects are found to affect the magnitude of the cross 
section.
Some possibilities reflecting the 
effects of the medium on the propagation and properties of the intermediate 
S$_{11}$ resonance are studied. 
The effects of allowing the S$_{11}$ to interact with the medium via 
mean field scalar and vector potentials are considered.
Both broadening of width and reduction in mass of the resonance lead to 
a suppression of the calculated cross sections. 
\end{abstract}
\hspace{-.25 in}PACS number(s)  25.20.Lj, 24.10.Jv, 14.20.Gk, 13.60.Le

\newpage
\section{Introduction}
\label{intro}
The photoproduction of the eta mesons on nucleons and nuclei is a subject of 
current interest. The production on nucleons provides an opportunity to study
nucleon resonances in the second resonance region. In the past decade there 
have been a large number of studies, both theoretical and experimental, on the
reaction on nucleons and deuterons and much has been 
learned \cite{ben,4TIA.PRC,fa1,KL}.

Photoproduction on nuclei can be valuable in learning about the changes in 
hadron properties in the nuclear medium. 
Inclusive photoproduction cross sections for a number of  nuclei
have been measured at MAMI by Roebig-Landau {\it et al.} \cite{RLPLB}
and recently at INS by Yorita {\it et al.} \cite{YORITA}; data on coherent
reactions are confined to the lightest nuclei. No data have been reported 
however on quasifree or incoherent photoproduction. Part of the reason 
for this is the smallness of the cross sections for these
processes \cite{fa2,plm1}. 
It is therefore imperative that efforts should continue 
in the direction of improving the theoretical calculation, both for their own
sake and also for better guidance to experiments.

In an earlier study we presented a relativistic model for
exclusive and inclusive photoproduction of eta mesons on 
nuclei \cite{moh1,moh2}. 
This
study was extended to incoherent photoproduction in ref. \cite{is}. The model
is based on an effective Lagrangian to describe the production mechanism
\cite{ben}. It includes contributions from nucleon
resonances in the second resonance region, from nucleon pole diagrams
and from vector meson diagrams. This approach has proved successful
in describing the experimental data for the elementary reaction. The other
key ingredients of the study on nuclei are the use of relativistic mean
field dynamics to treat the nucleon motion and to allow for final state
interactions of the outgoing particles with the residual nucleus.

In the earlier studies, two key approximations were applied to simplify
the calculation of the reaction amplitudes. The first of these was a
local approximation for the propagators. The other approximation was
to use a free (undressed) form for the propagators, in which the
interactions of the propagating particles with the nuclear medium were
ignored.

Non-locality effects in coherent photoproduction of pions and eta mesons
on nuclei have been studied, in a relativistic model, by Peters {\it et al.}
\cite{plm1,plm2}. For the pion production case these authors also investigated
the medium modifications of the nucleon and delta propagators. For the
coherent photoproduction of eta mesons it was established that non-local
effects can lead to enhancement of the contribution of the  $S_{11}$ 
resonance
which appeared to be strongly suppressed in earlier local calculations
\cite{fa2,psb}.

The present study is concerned with the investigation of 
 non-local
and medium effects in the quasifree photoproduction of eta mesons on
nuclei, i.e. reactions of the type $A(\gamma, \eta p)B$. In this sense it
complements the work of Peters {\it et al.}, on coherent photoproduction. 
Even though the cross section for the quasifree 
photoproduction 
reaction is rather small, it serves as a proto-type for studying 
these effects. Moreover its amplitude is the main building block, 
in at least some models \cite{moh2,4LEE.NPA}, for calculating 
inclusive cross sections. 

Earlier relativistic and non-relativistic studies of the quasifree
photoproduction \cite{moh1,4LEE.NPA} show clearly that the reaction is strongly
dominated, at the energies not too far from threshold, by the $S_{11}$ 
resonance. We therefore restrict
the present study to the dominant S$_{11}$ contributions.

In the next section we discuss the formalism for improving our previous model 
\cite{moh1} 
(henceforth referred to as I), to include both non-locality and 
possible medium effects on the resonance.
The results and discussions 
are given in
section  \ref{sec3} and the conclusion in section \ref{sec4}.

\section{Formalism}

In I, starting from a relativistic interaction 
Lagrangian for a system of photons, 
nucleons and mesons, we obtained a transition amplitude for the 
A($\gamma $, $\eta$p)A$-$1 reaction. 
To simplify the calculations a local approximation was adopted and free 
propagators 
were used for the intermediate resonances (see I for details). 
In the present treatment both approximations will be dropped. 
It is assumed here that the production of eta meson takes place through 
formation and 
subsequent decay of 
the $S_{11}$(1535) resonance.

\subsection{The Non-local Reaction Amplitude}
\label{sub2a}
 
At the tree-level the S-matrix for the A$(~\gamma,~\eta p~)$A-1 reaction, 
through the S$_{11}$ resonance, can be cast in the following from \cite{moh1},
\begin{eqnarray}
S_{fi} = &&\frac{e}{(2\pi)^{17/2}}\frac{\kappa^p_R  g_{{\eta} N R}}{ M+M_R}
{\left( \frac{M}{E_p} \frac{1}{2\omega_\eta} \frac{1}{2\omega_\gamma} 
\right)}^{1/2}
\sum_{J_B M_B}{\left( J_f, J_B; M_f, M_B| J_i, M_i \right)}
  \left [S_{J_i J_f} (J_B)\right]^\frac{1}{2}
\nonumber\\
\times&&
 \{\frac{}{} \int d^4 x d^4 y d^4 p \hspace{0.05 in} \bar{\psi}_{sf}(y)\phi_\eta(y)
\frac{ e^{-i p(y-x)}}{ { / \hspace{-0.095 in}p } - M_R + i\frac{\Gamma}{2} }
\gamma_5
{ { / \hspace{-0.105 in} k }_{\gamma} { / \hspace{-.09 in}
\epsilon } }\hspace{0.05 in}   e^{i k_{\gamma}
x} \psi_B(x) \nonumber
\\
&&+\int d^4 x d^4 y d^4 p  \hspace{0.05 in} \bar{\psi}_{sf}(y)
\gamma_5 { { / \hspace{-0.105 in} k }_{\gamma}
{ / \hspace{-0.09 in} \epsilon } }\hspace{0.05 in} e^{i k_{\gamma}y}
\frac{ e^{-i p(y-x)}}{ { / \hspace{-0.095 in}p } - M_R + i\frac{\Gamma}{2} }
\phi_\eta(x)\psi_B(x)\hspace{0.05
 in}\frac{}{}\},
\label{eq1}
\end{eqnarray}
where  $  S_{J_i J_f} (J_B) $ and 
$\left( J_f, J_B; M_f, M_B| J_i, M_i \right)$ 
are spectroscopic and 
Clebsh-Gordon coefficients, respectively. 
$M$, $M_R$ and $\Gamma$ are the nucleon mass, resonance mass
and  width, respectively. $E_P$, $\omega_\eta$ and $\omega_\gamma$ are the energies of
the outgoing proton, eta meson and incident photon, respectively. 
$\kappa^p_R $ and $g_{{\eta} N R}$ are the anomalous 
magnetic moment of the resonance and the coupling constant of 
the eta-nucleon-resonance vertex.
$\psi_{sf}(y)$, $\psi_{B}(y)$ and $\phi_\eta(y)$ are wave
functions of the outgoing proton, the bound (initial state) proton and the eta 
meson,
respectively.
The $\psi$'s are solutions of Dirac equations with appropriate mean field potentials 
and $\phi$ is a solution of the Klein - Gordon equation.
Instead of the local approximation used in I 
(this approximation is exact in the limit where both $\psi_{sf}$ and $\phi_{\eta}$ are plane waves), 
we perform a complete non-local 
calculation.

The integrals in (\ref{eq1}) can be computed as they stand, 
but it is possible to follow a somewhat simpler approach \cite{cooper}. 
We rewrite the two integrals in (\ref{eq1}) as,

\begin{equation}\label{eq3}
 {\cal I } = (2\pi)^{4}\int  d^4 y  \hspace{0.05 in} 
 \bar{\psi}_{sf}(y)
\phi_\eta(y)
W^s(y)
 +(2\pi)^{4} \int  d^4 y  \hspace{0.05 in} 
 \bar{\psi}_{sf}(y)
\gamma_5 { { / \hspace{-0.105 in} k }_{\gamma}
{ / \hspace{-0.09 in} \epsilon } }\hspace{0.05 in} e^{i k_{\gamma} y }
W^u(y),
\end{equation}
where
\begin{eqnarray}\label{eq4}
(2\pi)^{4}W^s(y) &=& \int d^4 x d^4 p
\frac{ e^{-i p(y-x)}}{ { / \hspace{-0.095 in}p } - M_R + i\frac{\Gamma}{2}}  
 \gamma_5
{ { / \hspace{-0.105 in} k }_{\gamma} { / \hspace{-.09 in}
\epsilon } }\hspace{0.05 in}   e^{i k_{\gamma}
x} \psi_B(x) \nonumber
\\
(2\pi)^{4} W^u(y) &=& \int d^4 x d^4 p
\frac{ e^{-i p(y-x)}} {{ / \hspace{-0.095 in}p } - M_R + i\frac{\Gamma}{2} } 
\phi_\eta(x)\psi_B(x), \label{ed32}
\end{eqnarray}
where the superindices  refer to s- and u-channel diagrams. 
If we act from the left with the operator
 ${ / \hspace{-0.095 in}p } - M_R + i\frac{\Gamma}{2}$ and then 
carry the integration over momentum, we obtain  the following 
Dirac-type linear differential equation,
\begin{equation}\label{eq5}
  ( \hspace{0.05 in}{ / \hspace{-0.095 in}p }-M_R + i\frac{\Gamma}{2} )W^i(y) = V^i(y),
\end{equation}
where the source terms for s- and u-channels are, respectively,
\begin{eqnarray}\label{eq6}
 V^s(y) &=& \gamma_5
{ { / \hspace{-0.105 in} k }_{\gamma} { / \hspace{-.09 in}
\epsilon } }\hspace{0.05 in}   e^{i k_{\gamma}
y} \psi_B(y) \nonumber
\\
 V^u(y) &=& \phi_\eta(y) \psi_B(y).\label{eq51}
\end{eqnarray}
After taking care of the time dependence, 
equation (\ref{eq5}) leads to the following second order differential equation for the space part of the upper
component of $W^{i}$ (which we denote $W^{i}_{up}({\bf r})$; note it  
continues to be spin dependent and similarly for $V^{i}({\bf r})$). 
The lower component of $W^{i}({\bf r})$ can be obtained from its upper component 
(see below).
\begin{equation}\label{eq7}
-{\bf p}^2 W^i_{up}({\bf r})+ \alpha \beta  W^i_{up}({\bf r}) = \beta V^i_{up}({\bf r})- {\bf \sigma} \cdot{\bf p} V^i_d({\bf r}), 
\end{equation}
where the indices up and d indicate the upper and lower components of the functions 
$ W^i $ and $ V^i $.
The $\alpha$'s and $\beta$'s are given by,
\begin{eqnarray}\label{eq8}
 \alpha^s& =& E_b + w_\gamma - M_R + i\frac{\Gamma}{2} \nonumber \\
 \beta^s & =& E_b + w_\gamma + M_R - i\frac{\Gamma}{2} \nonumber \\
 \alpha^u & =& E_b - w_\eta - M_R + i\frac{\Gamma}{2} \nonumber \\
 \beta^u  & =&E_b - w_\eta + M_R - i\frac{\Gamma}{2} 
,\label{eq70}
\end{eqnarray}
where $E_b $ is the energy of the bound nucleon.
The following expansions are used for $ W^i_{up}({\bf r}) $ and the source
part of equation (\ref{eq7}) (we drop 
the index $i$ for now),

\begin{eqnarray}\label{eq9}
 W_{up}({\bf r}) = \sum_{L, M, J} \frac{w^{M}_{LJ}(r)}{r} {\cal Y}^M_{L\frac{1}{2}J}(\Omega)
\nonumber \\
 V_{up}({\bf r}) - \beta {\bf \sigma} \cdot{\bf p} V_d({\bf r})=\sum_{L, M, J} 
\frac{\xi^{M}_{LJ}(r)}{r} {\cal Y}^M_{L\frac{1}{2}J}(\Omega).
\end{eqnarray}

Substituting the above expansions into 
 (\ref{eq7}) leads to the following second order radial differential equation,
\begin{equation}
 \left[\frac{d^2}{dr^2}-\frac{L(L+1)}{r^2}+\alpha\beta
 \right] w^{M}_{LJ}(r) = \xi^{M}_{LJ}(r).
\label{eq10}
\end{equation}
The presence of the resonance's width in equation 
(\ref{eq5}) makes both $\alpha$ and $\beta$ complex. 
The equation above, in the limit of 
no source, is 
similar to  the Bessel equation with a complex variable. 
The complexity of the 
argument  requires care in the choice of boundary conditions to insure proper asymptotic behavior.  
In the presence of sources the solutions  must be matched to  combinations of 
Bessel and Neumann type functions, which vanish at 
large distances.
 The above equation can be
 solved by a number of different methods. 
 The method which we adopt here is the Gauss
 elimination matrix \cite{cooper}. 
This method is suitable for solving  the differential equation (\ref{eq10}) 
with the boundary condition mentioned above.   

Substituting eq.(\ref{eq4}) into eq.(\ref{eq1}) and using partial wave expansions of all wave functions  \cite{jon},
the S-matrix 
in the distorted wave approximation (DWA) can be written as,
\begin{eqnarray}\label{eq11}
S_{fi} &=& \frac{ e }{\pi}\frac{1}{(4\pi)^\frac{1}{2}}
{\left( \frac{E_p + M}{E_p \omega_\eta \omega_\gamma} \right)}^{1/2}
\frac{\kappa^p_R  g_{{\eta} N R}}{ M+M_R}\delta( E_B 
+ \omega_\gamma - E_p  - \omega_\eta ) 
\nonumber \\
& &\times \sum_{ J_B M_B}{ \left( J_f, J_B; M_f, M_B| J_i, M_i \right)
                     { \left[ {\cal S}_{J_i J_f} (J_B) \right] }^{1/2}  } 
\nonumber \\
& &\times\sum_{LJM} i^{-L}Y^{M-s_f}_L(\hat{k}_f)
\left( L, 1/2 ; M-s_f, s_f| J,M\right)
\nonumber\\
& &\times\left\{ \sum_{L_\eta,M_\eta}\right.
 i^{{-L}_\eta}\left[Y^{M_\eta}_{L_\eta}(\hat{k}_\eta)\right]^*
\nonumber\\
& &\times\int d^3 y
\left[{\cal Y}^{M}_{L,1/2,J}(\Omega)\right]^*\left[f_{LJ}(r)
\hspace{.3cm}i \sigma\cdot \hat{\bf r}g_{LJ}(r)
 \right]
 v_{L_\eta}(r)Y_{L_\eta}^{M_\eta}(\Omega)
\left[\matrix{W^s_{up} \cr W^s_d} \right]
\nonumber\\
 & &\hspace{.1cm} +\sum_{L_\gamma} 
 \left[\frac{2L_\gamma+1}{4\pi}\right]^\frac{1}{2}
 i^{{L}_\gamma}
\nonumber\\
 & &\left.\times\int d^3 r
\left[{\cal Y}^{M}_{L,1/2,J}(\Omega)\right]^*\left[f_{LJ}(r)\hspace{.3cm}
i \sigma\cdot \hat{\bf r}g_{LJ}(r)
 \right]{j_{L\gamma}}(k_{\gamma}r)
 Y_{L_\gamma}^0(\Omega)\gamma_0\gamma_5
{ { / \hspace{-0.105 in} k }_{\gamma} { / \hspace{-.09 in}
\epsilon } }\hspace{0.05 in}\left[\matrix{W^u_{up} \cr W^u_d}
 \right]\right\},\nonumber\\
\end{eqnarray}
where $f_{LJ}(r)$, $g_{LJ}(r)$ are the upper and lower 
component radial wave functions of the outgoing nucleon, respectively. 
$v_{l_\eta}(r)$ describes the radial wave function of the eta meson.   
The expansions of the upper components of $W^i$ are given in equation 
(\ref{eq9}).
The lower components $W^i_d$ are related to the upper components by,
\begin{eqnarray}\label{eq12}
  W_d^i({\bf r}) = \frac{\sigma \cdot {\bf p} W^i_{up}({\bf r}) - V^i_d({\bf r})}{\beta_i}.  
\end{eqnarray}

\subsection{Inclusion of Medium Effects}
The properties of the resonances are expected to change in the nuclear medium. 
The two properties of interest are mass and width. It is often argued that the mass of 
the nucleon is  changed in the medium, and in the spirit of the  
Walecka model \cite{bsjw}, this could be effected by the scalar field. 
Thus our first attempt is to let the S$_{11}$ resonance interact with the nuclear medium 
through the nuclear scalar and vector  
potentials.
The formalism discussed in the previous subsection can be  modified to include 
these nuclear mean fields in the resonance propagator. We modify
 equation (\ref{eq5}) by 
 subtracting the vector potentials from the zeroth component of the resonance 4-momentum
 and adding the scalar potentials to its mass,
\begin{equation}\label{eq13}
  ( \hspace{0.05 in}{ / \hspace{-0.095 in}p }- \gamma^0U_v(y)-M_R -U_s(y) 
+ i\frac{\Gamma}{2} )W^i(y) = V^i(y).
\end{equation}
The differential equation obtained from (\ref{eq13}) is somewhat  
different from (\ref{eq10}); the parameters  
$\alpha$ and $\beta$ are now functions of $r$. The equation has the form,
\begin{eqnarray}\label{eq14}
\left[\frac{d^2}{dr^2}-\frac{l(l+1)}{r^2}+\alpha (r)\beta (r)
+(\kappa (l) + 1 )\frac{\beta^\prime (r)}{r\beta (r)}
+ \frac{\beta^{\prime\prime} (r)}{2\beta (r)}\right.
-\frac{3}{4}(\frac{\beta^\prime (r)}{\beta (r)})^2 
&&\left. \right] y_{lJ}^M(r) = \\ \nonumber
\sqrt{\beta ( r)}\tau_{lJ}^M(r) 
+ i\frac{\beta^\prime (r)}{\beta^\frac{3}{2} (r)}\zeta_{l^\prime J}^M(r)
- i\frac{1}{\beta^\frac{1}{2} (r)}\zeta^{\prime M}_{l^\prime J}(r)
+ i\frac{\kappa (l^\prime) + 1}{r\beta^\frac{1}{2} (r)}\zeta_{l^\prime J}^M(r),
&&
\end{eqnarray}
where $w_{lJ}^M(r) =\sqrt{\beta ( r)} y_{lJ}^M(r) $ and the function $\kappa(l)$ 
is defined as
 \begin{eqnarray}\label{eq15}
 \kappa (l)& =& l +1 \hspace{1 cm} \mbox{for} \hspace{1 cm} J = l +\frac{1}{2}
 \\
\nonumber \
\kappa (l)& =& -l -1 \hspace{1 cm} \mbox{for} \hspace{1 cm} J = l -\frac{1}{2},
\end{eqnarray}
and $l^\prime = 2J -l$.

The inclusion of the interaction  
with the medium also necessitates that the expansion of the source functions 
( eq.(\ref{eq9})) be done for the upper and lower components separately,
\begin{eqnarray}\label{eq16}
 V_{up}({\bf r}) = \sum_{l, M, J} \frac{\tau^{M}_{lJ}(r)}{r} {\cal Y}^M_{l\frac{1}{2}J}(\Omega)
\nonumber \\
 V_d({\bf r}) =\sum_{l, M, J} \frac{\zeta^{M}_{lJ}(r)}{r} {\cal Y}^M_{l\frac{1}{2}J}(\Omega).
\end{eqnarray}
The second order radial differential equation (\ref{eq14}) is again solved 
using 
the gauss elimination matrix method mentioned above. The lower components of 
the $W^i$ are obtained 
from their upper components using equation (\ref{eq12}). 
The rest of the calculations 
proceed along the same lines as the non-locality ones of subsection \ref{sub2a}. 

The influence of the medium may also be formulated in terms of changes in the resonance 
properties such as mass and width. 
These changes are likely to be density dependent. When this is the case, these 
changes can be accomodated in the calculation of the reaction amplitude in 
essentially the same manner as the above treatment of interaction potentials. 
The implementation of non-density-dependent changes is of course much simpler.

\section{results and discussion}
\label{sec3}
In this section we investigate the impact of doing non-local
calculations on the cross section for the quasifree
production. We also consider a number of
options that would simulate the medium effects on the
propagating S$_{11}$ resonance and asses how these affect the calculated cross 
sections. 

In I we have investigated the kinematical conditions under which the
calculated cross sections are optimal.
It was concluded that a symmetric arrangement for detecting
the outgoing proton and eta meson at 30$^o$ on both sides of
the incident beam led to the maximum cross section on $^{12}$C at
750 MeV. We shall adopt this geometry in all the
calculations presented here. In addition the coupling
parameters used are the same as those of I. 
The bound state wave functions for the
initial state bound proton are calculated using the mean
field Hartree potentials of Horowitz and Serot \cite{4HOR.NPA}. 
The distorted wave functions
for the outgoing protons are calculated using golobal potentials 
provided by Cooper {\it et al.} \cite{cooper1} and 
the optical potentials for eta mesons are those of 
Chiang  {\it et al.} \cite{4OSE.PRC} with 
the real part of S$_{11}$ self energy being taken to be prportional to 
the nuclear density.

Figure 1 shows the calculated triple differential cross
section at E$_\gamma$ = 750 MeV as a function of the kinetic
 energy of the
outgoing eta meson. As pointed out earlier, a calculation in
which both the proton and the eta are taken as plane waves is
by necessity a local calculation. We show this by the dotted
curve. The dashed curve shows  the corresponding local
distorted wave (DW) calculation. Comparison between these
two curves establishes, as has been learned in many earlier
calculations, the strong suppression of the cross sections
due to final state interactions. 

The solid curve in Fig.1 shows the non-local calculations.
Here we find that the non-locality effects lead to increase
in the cross section, but with only a slight change in
shape. The increase can be as high as 25\%. Figure 2 shows a
similar comparison between local (dashed curve) and
non-local calculations (solid curve) for $^{ 40}$Ca at the same incident
energy. The features are essentially the same as in Fig.1.
Thus the increase in the cross section due to non-locality
effects appears to be independent of the target nucleus.

Some theoretical approaches to the calculation of the
inclusive cross sections start from calculations of the
exclusive cross sections discussed here but with the wave
functions of the outgoing protons taken as plane waves. Thus
in these calculations only the eta wave function is
distorted due to the final state interactions. The rationale
for this is given, for example in ref. \cite{moh2}. In
Fig.3 we show the effect of non-locality on this type of
calculation. We see again that the net result is an increase
in the cross section in essentially the same fashion as in the
two cases discussed in Figs.1 and 2.

The increase of the cross section in non-local calculations
appears to be a universal feature independent of the energy
of the incident photon (in the region of the S$_{11}$ dominance). It is
also present at other angular pairs of the outgoing
particles provided these remain in the forward hemisphere
close to where the cross section is large. It is of interest
to try to understand why the inclusion of non-locality leads
to an increase in the cross sections. This feature appears
to be closely connected to the final state interactions of
the outgoing particles with the residual nucleus, i.e. to
the distortion effects. To see this we first note that plane
wave calculations in which these effects are neglected are
strictly local and as seen in Fig.1 have relatively large
cross sections. Local calculations with distortion effects
lead to a noticeable reduction of the cross section. But the
cross section increases as the non-locality is taken into
consideration.

Because the cross sections are substantial only when the
pair of outgoing particles are moving in a forward direction
it is possible to argue that the increase is due to a
reduction in the absorption of the outgoing particles in the
non-local case. In the plane wave local calculations the
pair is produced at the same point where the formation of
the S$_{11}$ resonance takes place; subsequently they move freely
through the nuclear medium. In local distorted wave
calculations they do interact with the nucleus and suffer
absorption and hence the cross section is reduced (see
dashed curve in Fig.1). In non-local calculations, on the
other hand, the S$_{11}$ resonance decays generally at a point
different from the point of its formation. Because the
motion is largely forward the outgoing particles travel in
the nuclear medium for a shorter distance and hence suffer
less absorption. This leads to some increase in the cross
section. We have tested this interplay between absorption
effects and non-local calculations by gradually scaling the
distorting potentials of either particle. We found clear confirmation of this
relationship between the relative increase in the cross
section and the size of the distorting potentials. Note that
the reason for the increase in the present quasifree
reaction is quite different from the case of coherent
photoproduction where the increase in the cross section due
to non-locality can be understood in terms of the interplay
between the nuclear structure of the target and the spin
structure of the elementary amplitude \cite{plm2}.

We now investigate the influence on the cross section due to
the medium effects on the propagating S$_{11}$ resonance. 
All the calculations reported below include
non-local effects as discussed above. It is generally agreed
that the medium is likely to affect both the mass and the
width of the resonance. We can alternatively look at these
effects in a framework where they are expressed in terms of
interactions of the resonance with the surrounding nucleons. In
this respect one is inclined to think of a mean field
interaction modeled along the case of nucleons. It may
therefore be a good starting point to consider such a case.
A rough approximation is to take the interaction to be that
experienced by the bound nucleon that formed the resonance,
namely the scalar and vector Dirac Hartree potentials of ref.
\cite {4HOR.NPA}. The results are displayed in Fig.4. 
The solid curve (No ME)  represents the case of free propagation,  
i.e without
interactions with the medium. If we include only the scalar
Hartree potential S(r) in the calculations, we are then
affectively changing the mass of the resonance in
radially-dependent way.  The result of invoking such an
effective mass is shown by the dashed curve in the Fig.4.
There is a large reduction in the cross section. 
We should add however that the
Hartree scalar potential is rather deep (about 600 MeV at
the centre of the nucleus) and hence the effect may be
somewhat exaggerated. The dotted curve shows the results
when both the Hartree vector and scalar potentials are
included. We find, in this case, a much reduced effect on the
cross section; the  vector and scalar potentials have
opposing influences. This can be understood qualitatively in
terms of the opposite signs of these potentials and the
structure of the propagator. The vector potential when added
to the energy term tends to offset the increased gap between
the energy and mass terms resulting from the reduction in
mass due to the scalar potential.

Another possible way of probing the effects of the medium on 
the propagating
resonance is to change the width of the resonance from its
free space value. The calculations discussed above have all
been performed using $\Gamma$ = 150 MeV. In addition to
uncertainties surrounding the free width, it  is likely that
the width will broaden in the medium; there are for example 
indications of some broadening 
in the region of the $\triangle$(1232) resonance \cite{EM}. 
To test the effect of such broadening we carry out
calculations using a larger value   $\Gamma$  = 208 MeV. 
This value is obtained by 
Breit - Wigner fits to the data of the elementary reaction 
\cite{BKPRL} and is also close to the  upper limit used 
in the analysis of ref. \cite{YORITA}.
We also consider
the effect of collision broadening of the width following a
suggestion by Lehr and Mosel \cite{LM}. These authors considered
a model for the broadening of the D$_{13}$ resonance due to
collisions with nucleons, in an attempt to explain the total
photoabsorption cross section on nuclei. They suggest that,
to first order in the baryon density $\rho$,  the width may be
written as 
\begin{eqnarray}\label{eq17}
\Gamma( \rho ) = \Gamma_{\mbox{free}} + 50 \frac{\rho}{\rho^0}
\end{eqnarray}
where $\Gamma_{\mbox{free}}$ = 150 MeV and $\rho^0$ is the density 
of infinite nuclear matter (taken to be 0.17 fm$^{-3}$).

The results of these calculations are shown in Fig.5. We
note that increasing the width from 150 to 208 MeV results
in a drastic reduction (about 50\% at the peaks) of the
cross section. On the other hand the density dependent
broadening has only a small effect (around 10\%). The
weakening of broadening in the surface region in this case
appears to be responsible for the much reduced effect on the
cross section.

The next step in looking at medium effects is to reconsider
the changes in the mass of the resonance in the medium
together with the change in width. Figure 6 shows comparisons
involving the use of an expression for the mass borrowed
from the work of Saito and Thomas \cite{ST} on the medium
effects on baryon masses. One of the relations used by
these authors for the change in the nucleon mass within the
medium is of the form
\begin{eqnarray}\label{eq18}
M( \rho ) = M_{\mbox{free}} - 0.14  \frac{\rho}{\rho^0} M_{\mbox{free}}
\end{eqnarray}
We adopt this same expression for the mass of the S$_{11}$. 
This provides a somewhat different form for the
change in mass from that used in the calculations of Fig.4 (due the scalar
part of the Hartree potential). The cross section calculated
with the form (\ref{eq18}) for the mass is shown by the dashed
curve in Fig.6. The solid curve is the result of
calculations that include the above mass change as well as a
change in width according to the form given in eq.(\ref{eq17}). The
mass dependence on the density reduces the cross section by
an amount which, though substantial, is much less than what
was obtained using the deep Hartree potential (Fig.4). The addition
of density-dependent width broadening reduces the cross
section further, similar to what was noted in Fig.5.

Finally we discuss the medium effects for the calculations
shown in Fig.3, where only the final state interactions of
the eta are taken into account. The comparisons shown in
Fig.7 are for the density dependent mass and width changes
discussed above. We see that the mass effect is larger than
the width effect and that the two together lead to the
smallest cross section. It is therefore likely that the net
effect of the medium on inclusive reaction will be to reduce
the cross sections. Detailed calculations of this are underway. 

We have carried out similar calculations at E$_\gamma$ = 900 MeV. 
The general features observed in the above figures are found to hold 
at this higher enery.

\section{conclusion}
\label{sec4}
Our earlier work on the exclusive quasifree photoproduction
of eta mesons on nuclei involved two approximations in the
calculations of the production amplitude. The first was a
local approximation in which the absorption of the photon
and the production of the eta were assumed to occur at the
same point. The second approximation was to assume free
propagation of the intermediate resonances; no interaction
or change of properties took place in the medium. In this
paper, adopting an S$_{11}$ resonance model, we have carried out
non-local calculations in order to correct for the first
approximation. We have also investigated the role of medium
interactions or alternatively changes in mass and width of
the resonance on the calculated cross sections. 

 We find that the non-local effects are important for the
cross section calculations. Typically these effects lead to
a moderate increase in the cross sections. This feature
appears to apply regardless of the target nucleus or whether
the final state interactions of the proton are taken into
account. 

In our investigation of the influence of possible medium
effects on the propagation of the S$_{11}$ resonance we
considered various possibilities. If one assumes that the
resonance experiences a mean field similar to that of the
nucleon then the scalar part of the mean field would act as
a de-facto reduction in the mass of the resonances. We find
that this term alone would lead to large suppression in the
cross section if its strength is maintained at that used for
nucleon binding in a Dirac - Hartree model. This suppression
is found to moderate if the vector component of the
interaction is added. Note however that the Hartree
potentials are known to be rather deep.
   
The cross sections are no doubt sensitive to the mass of the
resonance. In fact we found that if the mass of the
resonance is reduced by 30\%, the cross section is
drastically reduced. A more realistic change in the mass is
that of ref. \cite{ST},  based on the quark meson coupling model. In
this instance also the mass is dependent on the nuclear
density but in a    much weaker fashion compared to that
affected by the Hartree scalar potential. The cross section
is reduced due to the reduction in mass; the moderate
density dependence leads to a moderation in the suppression
of the cross section. Increasing the width of the resonance
also leads to a reduction in the cross section. If the width
broadening is density dependent the suppression of the cross
section is reduced.  

The effects noted above carry over to the case where only
the final state interactions of the eta meson are taken into
account. This latter case is relevant to the calculation of
the inclusive cross sections on nuclei.

In conclusion we have shown the importance of including
non-local effects in the relativistic calculations of the
cross section for quasifree exclusive photoproduction of eta
mesons on nuclei. In this framework this allowed us to probe
the dependence of the cross section on interactions with the
medium or alternatively on changes in the mass and width of
the S$_{11}$ resonance. This may open the way to using our
present model to investigate these medium effects in
relation to data on the inclusive reaction on nuclei (note
that the cross sections for the exclusive reaction are
rather small). Currently data that extend over a useful
photon energy range exist only for the $^{12}$C nucleus. We are
in the process of analyzing these data. However it is
desirable that the range of energies for existing data on
other nuclei be extended to allow their inclusion in such
analyses.
\section*{Acknowledgements}
Two of the authers (M.H. and S.B.) whish to acknowledge the support of 
Faculty of Science  Tehran University.

\newpage
                
\section* {Figure Captions}
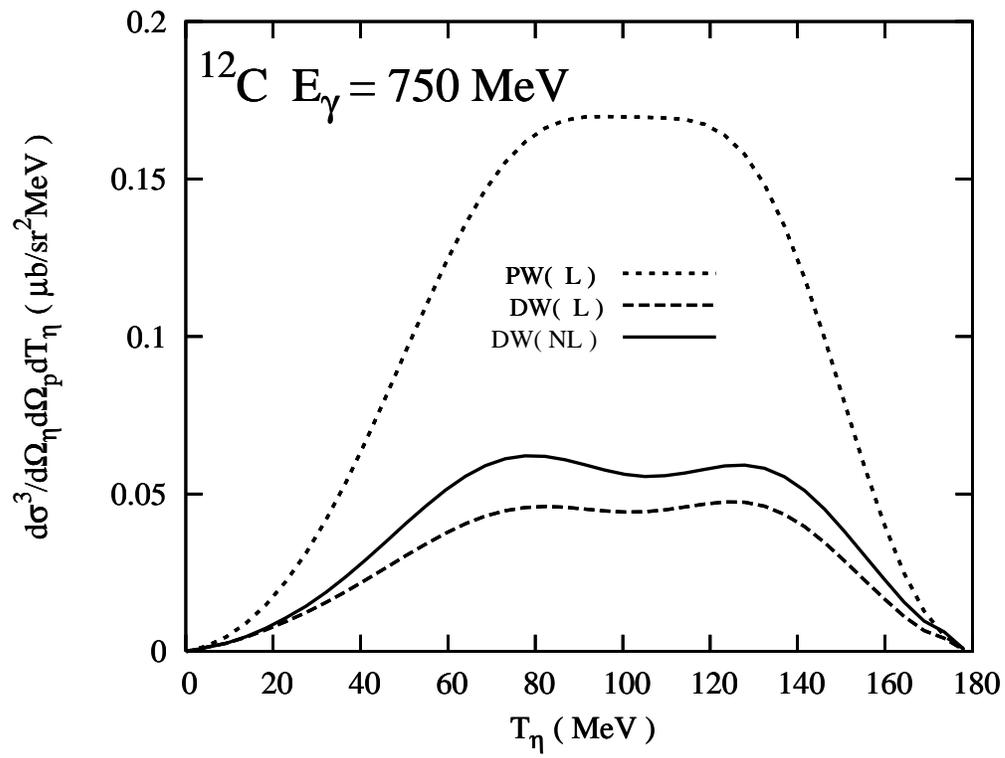
\noindent FIG. 1.  The cross section for the 
$^{12}C(\gamma, \eta p)^{11}B_{g.s.}$ 
reaction at photon energy of 750 MeV, 
plotted as a function of the kinetic energy of the outgoing $\eta$ meson. 
Solid curve : non-local DW calculations (curve labeled DW(NL)).
Dashed curve : local DW calculations (curve labeled DW(L)).
Dotted curve : local PW calculations (curve labeled PW(L)). 
In this and all the subsequent figures S$_{11}$ dominance is assumed.

\vspace{2.5 mm}
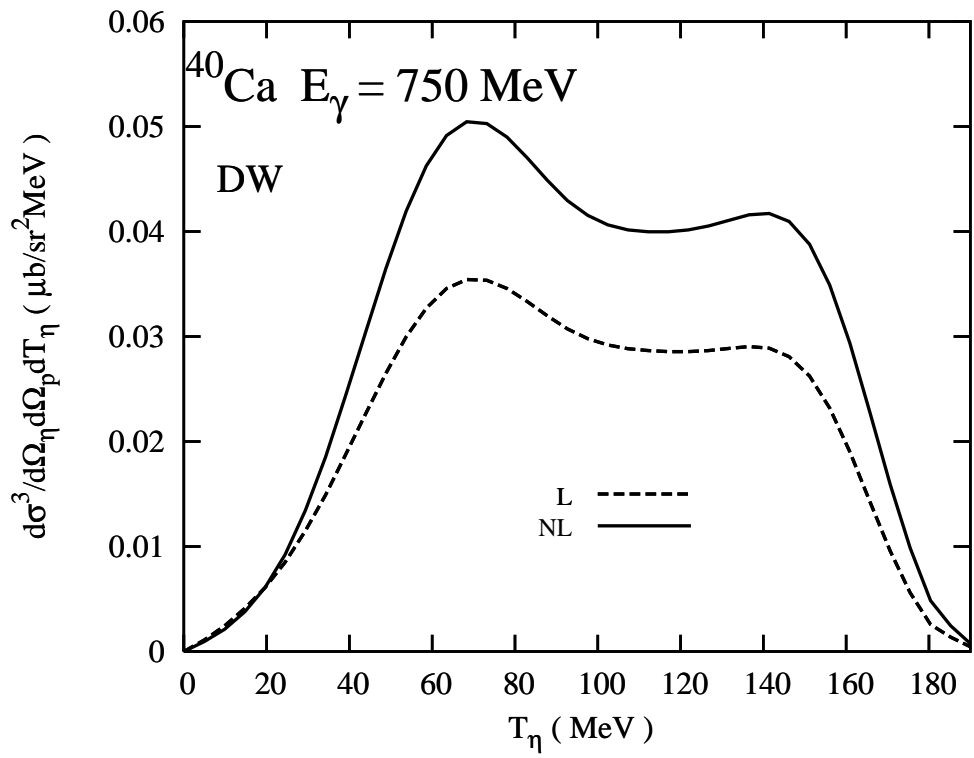
\noindent FIG. 2.  The cross section for the 
$^{40}Ca(\gamma, \eta p)^{39}K_{g.s.}$ 
reaction at 750 MeV. 
Only DW calculations are presented;
solid curve : non-local DW calculations;
dashed curve : local DW calculations.

\vspace{2.5 mm}
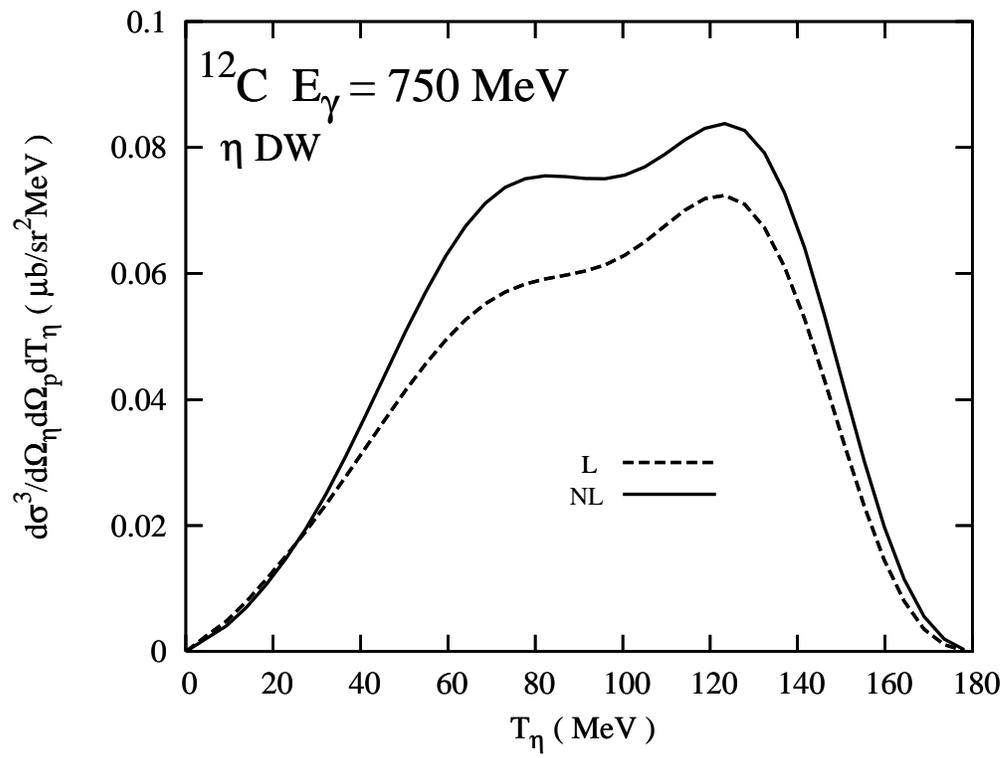
\noindent FIG. 3.  Same reaction as Fig.1. 
Distorted waves are used for the etas but protons are described by plane 
waves. 
Solid curve : non-local calculations;
dashed curve : local calculations.

\vspace{2.5 mm}
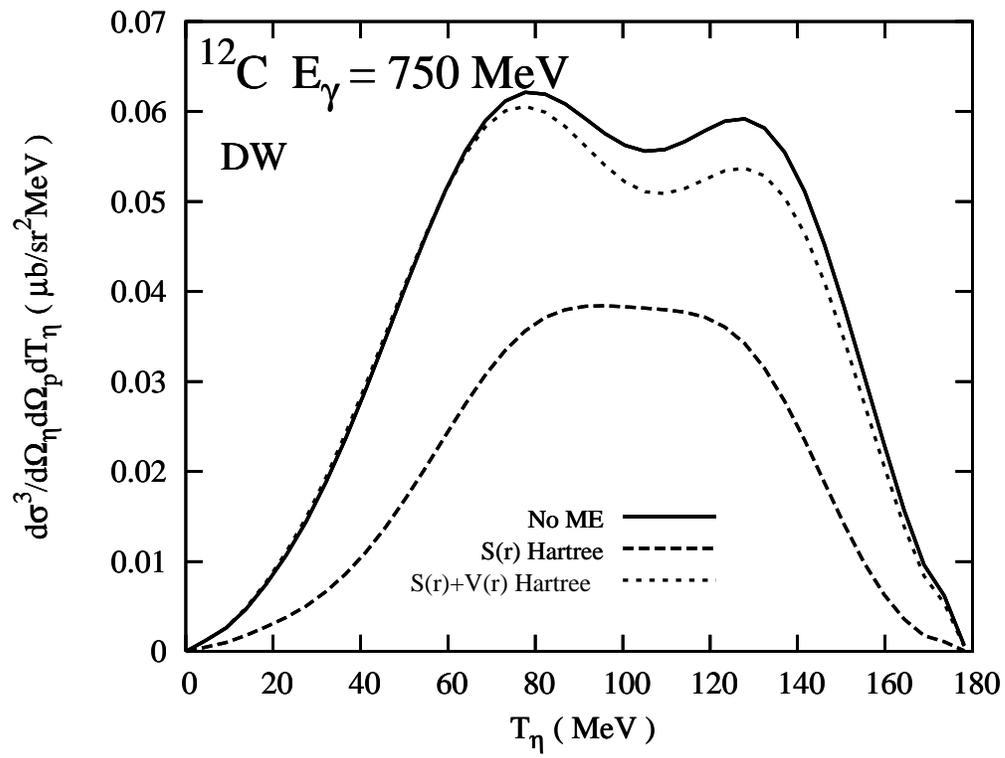
\noindent FIG. 4.  Same reaction as Fig.1. 
Solid curve : calculations using a free propagator for S$_{11}$ 
(curve labeled No ME).
Dashed curve : the resonance interacts with medium via Hartree scalar 
potential (curve labeled S(r) Hartree).
Dotted curve : the resonance interacts with medium via Hartree scalar 
and vector potentials (curve labeled S(r)+V(r) Hartree).

\vspace{2.5 mm}
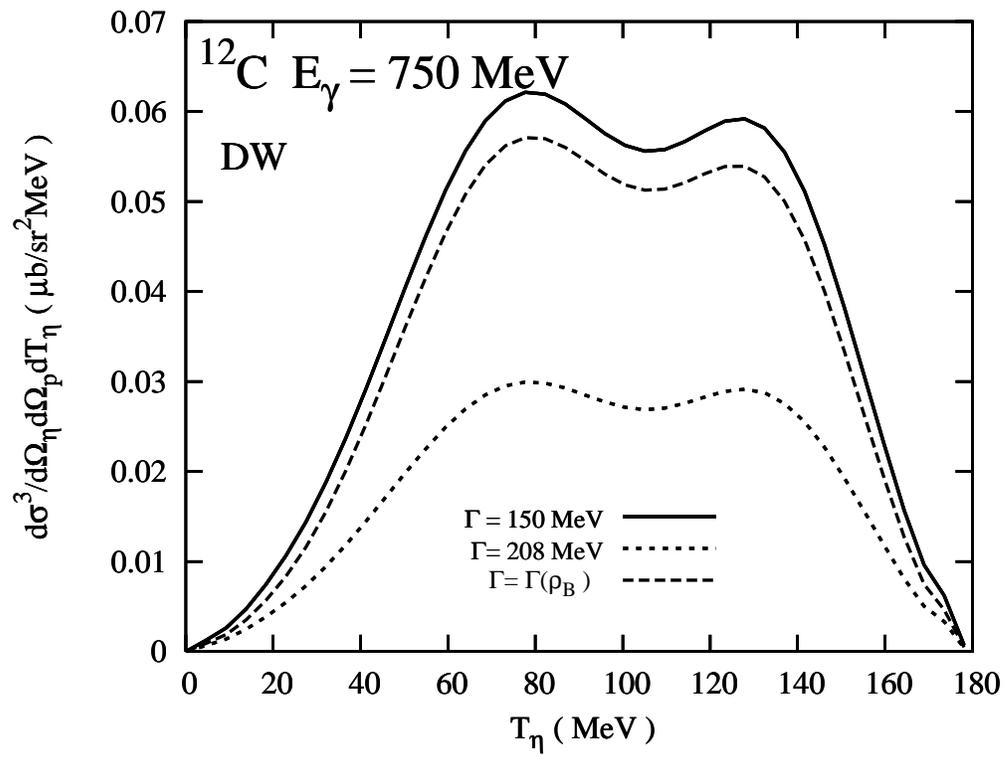
\noindent FIG. 5.  Same reaction as Fig.1. 
Solid curve : calculations using a free propagator with width $\Gamma$ = 150 MeV for S$_{11}$; 
dotted curve : $\Gamma$ = 208 MeV.
Dashed curve : calculations with a density dependent width (see text).

\vspace{2.5 mm}
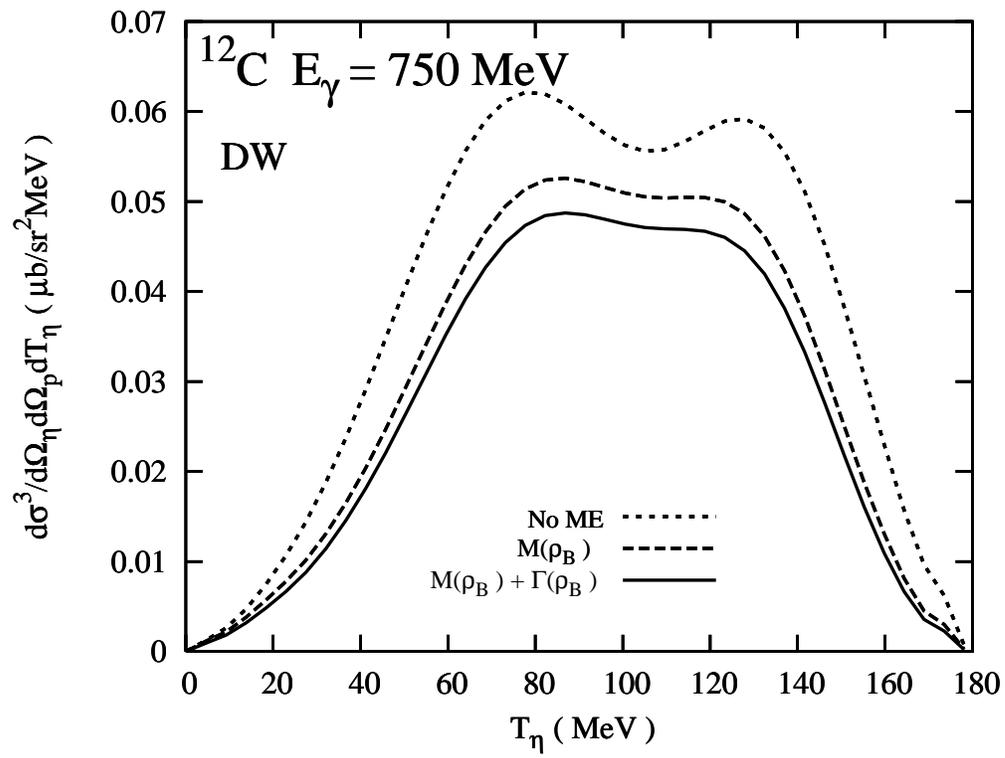
\noindent FIG. 6.  Same reaction as Fig.1. 
Dotted curve : calculations using a free propagator for S$_{11}$  
(curve labeled No ME).
Dashed curve : calculations with density dependent mass for S$_{11}$  
(curve labeled M($\rho_B$)).
Solid curve : calculations including density dependence for both mass and width of S$_{11}$.  

\vspace{2.5 mm}
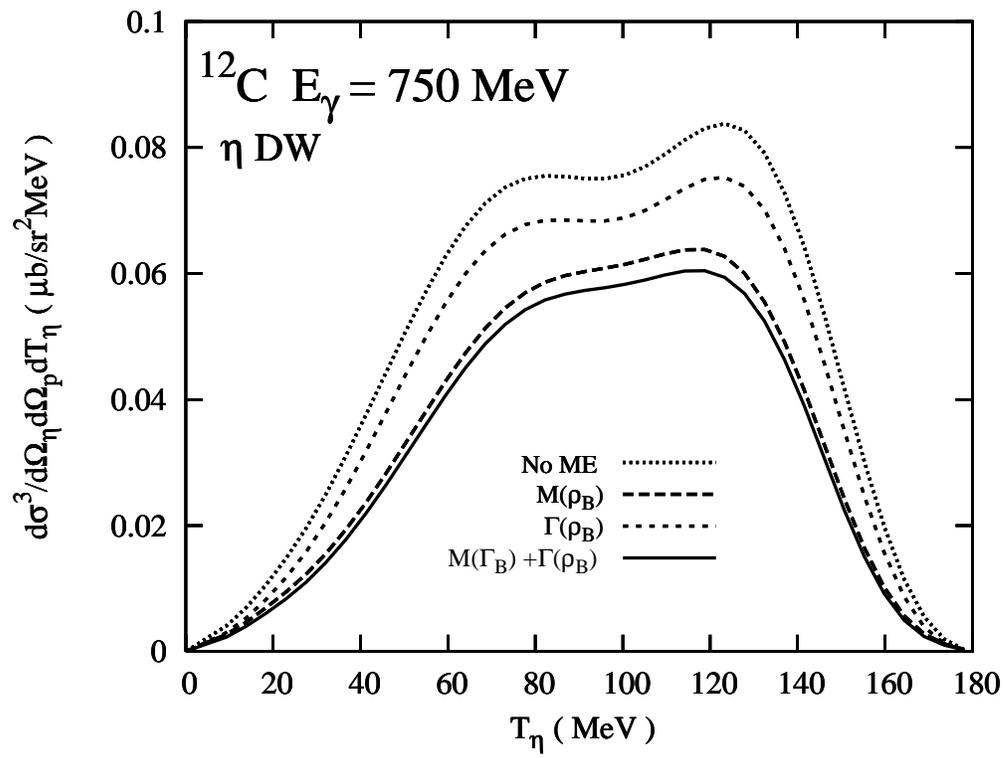
\noindent FIG. 7.  Same reaction as Fig.3. 
Dotted curve : calculations using a free propagator for  S$_{11}$.
Long-dashed curve :  calculations with density dependent mass for S$_{11}$.    
Short-dashed curve :  calculations with density dependent width for S$_{11}$.  
Solid curve : calculations include density dependence for both mass and width of S$_{11}$.

\vspace{2.5 mm}
\end{document}